# Colorectal cancer trends in Chile: a Latin-American country with marked socioeconomic inequities

Susana Mondschein*[1,2], Felipe Subiabre[1], Natalia Yankovic[3], Camila Estay[4], Christian Von Mühlenbrock[4,5], Zoltan Berger[4]

(1) Industrial Engineering Department, Universidad de Chile, Santiago, Chile.

(2) Instituto Sistemas Complejos de Ingeniería, Santiago, Chile

(3) Universidad de los Andes, Chile. ESE Business School.

(4) Department of Medicine, Gastroenterology Section, Hospital Clínico de la Universidad de Chile.

(5) Internal Medicine Department, Universidad de los Andes.

*Corresponding author: susana.mondschein@uchile.cl (SM)

**Abstract**

**INTRODUCTION**: Colorectal cancer (CRC) is the third most frequent malignant disease in the world. In some countries with established screening programs, its incidence and mortality have decreased, and survival has improved.

**AIMS:** To obtain reliable data about the epidemiology of CRC in Chile, we analyzed the trends in the last ten years and the influence of observable factors on survival, including explicit guarantees in CRC treatment access (GES program).

**METHODS**: Publicly available data published by the Health Ministry and National Institute of Statistics were used. Data were obtained from registries of mortality and hospital discharges, making follow-up of the individuals possible. Crude and age-standardized incidence and mortality rates were calculated, and individual survival was studied by constructing Kaplan–Meier curves. Finally, a Cox statistical model was established to estimate the impact of the observable factors.

**RESULTS:** Ninety-nine thousand and eight hundred forty-six hospital discharges were registered between 2009 and 2018 in Chile, corresponding to 36,649 patients. In the same period, 24,154 people died of CRC. A nearly linear, steady increase in crude incidence, mortality and prevalence was observed. CRC incidence was the lowest in the North of the country, increasing toward the South and reaching a maximum value of 35.7/100,000 inhabitants/year in terms of crude incidence and 20.7/100,000 inhabitants/year in terms of crude mortality in the XII region. Kaplan–Meier survival curves showed a slight improvement during the study period. The survival was shorter in people older than 70 years, but without significant differences in the younger age groups. Depending on socioeconomic status, survival was significantly better with private insurance than the national insurance system. Patients in the capital city survived longer than those in other parts of the country. We found no significant effect on survival associated with the GES program.

**CONCLUSIONS:** The introduction of a national screening program with rapid access to diagnostic and therapeutic procedures is the only way to diminish serious inequality and improve the survival rate of CRC in Chile.

# 1. Introduction

Colorectal cancer (CRC) represents up to 10% of cancers diagnosed worldwide each year and is the second and third most common in women and men, respectively (1). CRC is related to lifestyle, and its incidence is increasing around the world (2). Incidence rates vary according to geographic area, with the highest levels in developed countries and lower levels in developing countries (3). It is estimated that by 2035, there will be 2.5 million new cases diagnosed, and the incidence of CRC in Latin America by 2030 will increase by 60%, with a total of 396,000 new cases per year (4). The total number of deaths attributed to CRC is projected to increase by 60% and 71.5% in the colon and rectum, respectively, between 2013 and the projection for 2035, in part due to population growth and aging (5).

Reliable data on the statistics of CRC in Chile are relatively scarce, and most of the national investigations have focused on mortality rates, all of which have shown an upward trend over the years (6) (7) (8). In fact, the latest publication reports that crude mortality for CRC in Chile for 2016 was 9.18 per 100,000 people, increasing more than 20% between 2000 and 2016 (8).

There are no publications regarding CRC incidence and survival rates. Official statistics (DEIS, MINSAL) precisely register the number of hospital discharges, but there is no individualized number of cases with first diagnosis of CRC. Chile has a particular geography with nearly 5,000 km longitudinal extension with varying climatic conditions, nutritional habits, and ethnic composition. Epidemiological differences have been reported in gastric and gallbladder cancer, showing a higher frequency in *the Mapuche* population, concentrated in southern Chile (9) (10) (11).

The most impressive statistic is that the mortality/incidence ratio is twice that in Latin America and the Caribbean, where six out of ten patients die, while in the USA, it is three out of ten (12). Some plausible explanations refer to our different health care systems and early screening strategies, which are advanced in the USA and practically absent in Latin America. Notably, extreme inequality in social factors such as education and income has caused poor outcomes in cancer survival. Moreover, the personal and familiar economic consequences of a CRC diagnosis strongly depend on health insurance.

Several publications (13) (14) (15) are dedicated to analyzing the survival of CRC in different specialized centers, depending on the phase of the disease. However, we have no information on the lethality of CRC at the national level. Survival depends on the diagnosis of the disease in the earliest stage possible and on rapid access to adequate treatment.

In Chile, there is no screening program at the national level. The first screening program implemented in our country was known as PREVICOLON, a prospective, multicenter study conducted between 2007 and 2009, followed by the PRENEC (Prevention of Colorectal Neoplasms) program, a Chilean-Japanese collaboration, with promising results on early CRC diagnosis (16) (17). Regarding treatment access, the GES program (GES: explicit guarantees in health - a set of guarantees aimed to ensure prompt access to affordable, and quality health care) includes CRC from 2014 and covers diagnostic and therapeutic procedures from the moment of suspected CRC (18).

To determine a median-to-long term strategy at the national level, it is necessary to know the real situation of the disease. The primary aim of our present study was to obtain reliable information on annually diagnosed new CRC cases in Chile, i.e., the incidence of CRC, and to describe their epidemiological characteristics and geographical distribution. The secondary aim was to analyze the influence of several observable factors on patient survival, including some socioeconomic aspects, namely, differences in health insurance and the introduction of the GES program.

## 2. Materials & Methods

The Chilean health care system is a hybrid of public and private providers and insurances consisting of i) Fondo Nacional de Salud (FONASA – National Health Fund), which is public insurance for 78% of the Chilean population; ii) private health care insurers (ISAPREs) for 14% of the population; and iii) the Military and Police Forces' health system that represents 2.8% of the population. Privately insured patients can only access private providers (with a variety of coverages), while FONASA patients – paying a lower monthly fee – may access public and private providers depending on their income level with different copays.

There are significant socioeconomic differences among people belonging to each health care system. For example, in the first decile (poorest), FONASA represents 92% and ISAPRE represents 2%, compared to the tenth decile (richest), where FONASA represents 25% and ISAPRE represents 68% of the population (19). Appendix A characterizes the subgroups for FONASA insurance.

### 2.1 Data description

We used the national registry of all inpatient discharges from hospitals in Chile, considering both the public and private sectors, for the period between 2001 and 2019. The database has 39 fields, including primary and secondary diagnosis, sex, age, ethnicity, health insurance, hospital, region of residency, length of stay and condition at discharge.

We constructed a treatment database considering all patients for whom we had a first registry between 2009 and 2018 with a diagnosis code associated with CRC, resulting in 36,649 patients, corresponding to 99,846 different hospitals discharge episodes with a mean of 2.7 (std 4.03) hospitalizations per patient.

Using the national death registry, we constructed a mortality database considering all deaths between 2009 and 2018 for patients with a primary diagnosis of CRC. The constructed death database has 24,154 patients. This database includes 6,626 patients who are not in the treatment database (without any CRC principal or related hospital discharge between 2009 and 2018). Table 1 presents the characteristics of the constructed databases for the period under study.

|  | 2009 | 2010 | 2011 | 2012 | 2013 | 2014 | 2015 | 2016 | 2017 | 2018 | TOTAL |
|---|---|---|---|---|---|---|---|---|---|---|---|
| New identified CRC patients | 2,909 | 2,886 | 3,141 | 3,332 | 3,381 | 3,718 | 3,991 | 4,155 | 4,521 | 4,615 | 36,649 |
| Average Age (std) | 66.3 (14.0) | 66.5 (14.1) | 66.0 (14.1) | 65.8 (13.9) | 65.7 (13.9) | 66.1 (14.0) | 66.2 (13.6) | 66.1 (13.3) | 65.6 (13.6) | 66.1 (13.5) | 66.0 (13.8) |

| Women (%) | 51.2 | 52.0 | 51.5 | 51.6 | 50.3 | 49.5 | 50.5 | 49.8 | 49.2 | 49.0 | 50.3 |
|---|---|---|---|---|---|---|---|---|---|---|---|
| Total CRC deaths | 1,906 | 1,952 | 2,121 | 2,239 | 2,375 | 2,511 | 2,663 | 2,691 | 2,773 | 2,923 | 24,154 |
| Not in treatment database (*) | 547 | 554 | 576 | 582 | 655 | 719 | 725 | 783 | 727 | 758 | 6,626 |
| Average Age (SD) | 71.5 (13.8) | 71.5 (13.8) | 71.6 (13.1) | 71.9 (13.2) | 70.9 (14.0) | 71.9 (13.5) | 71.7 (13.8) | 72.0 (13.6) | 71.6 (14.0) | 71.7 (13.7) | 71.6 (13.7) |
| Women (%) | 54.1 | 50.5 | 52.5 | 52.3 | 52.1 | 50.9 | 51.0 | 51.5 | 50.8 | 49.2 | 51.4 |

*Table 1. Data description. The first part of the table characterizes the treatment database, while the second part characterizes colorectal cancer deaths. (*) Represents patients only appearing in the death registry without any CRC-associated discharge registry. The number of total CRC deaths also included these patients.*

Inclusion and exclusion criteria

For both the treatment and death databases, the primary and secondary diagnoses were encoded using the International Classification of Diseases, 10th Revision (ICD-10) codes.

We therefore identified two relevant subsets of the ICD-10 codes (see Appendix B). The first one, called principal codes, includes those directly identifiable as corresponding to colorectal cancer. The second group, called related codes, includes ICD-10 codes that do not by themselves indicate a CRC diagnosis but can be confidently linked to it when paired with a principal code. These are relevant to correctly distinguish the patients' causes of death (whether by CRC or an unrelated cause) and identify CRC survival times.

We also mention that within the principal codes are those corresponding to benign tumors, which are not considered relevant when appearing in the treatment database but are relevant when used as a cause of death. Figure 1 summarizes the inclusion and exclusion criteria for the treatment and death databases.

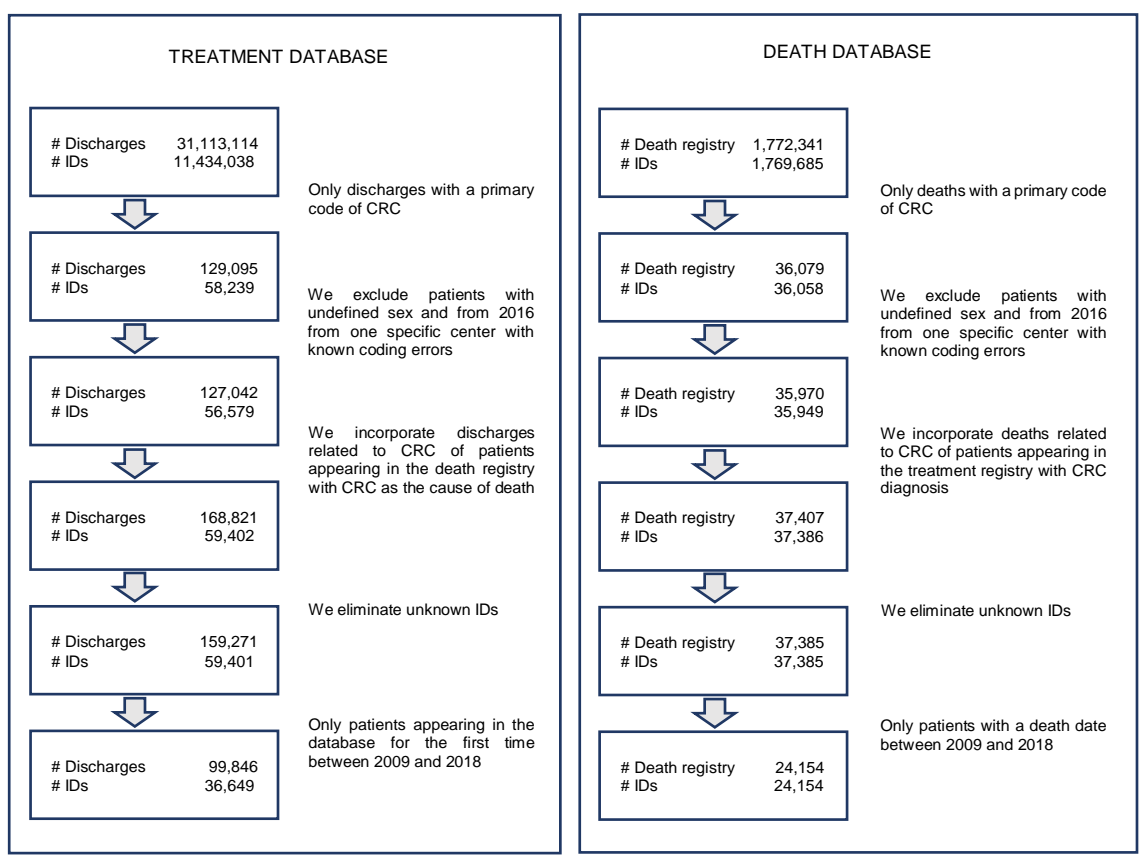

Figure 1. Construction of treatment and death databases – inclusion and exclusion criteria. Source: DEIS patient discharge database 2001-2019, DEIS mortality database 2000-2018. Unknown and unavailable IDs were eliminated.

## 2.2 Methods

We used publicly available data at the Ministry of Health. All data are protected, and personal information is anonymized. Crude- and age-standardized incidence and mortality rates were computed for the total population using the Segi World standard population table (20).

For each patient in the treatment database, we built a set of possible predictors for their survival. They include sex, health insurance, region of residency, age at diagnosis, year of diagnosis and coverage by the GES guarantees.

This last condition is determined by the year of treatment (GES was incorporated in 2014, thus covering half of our study period) and health insurance, since patients in the Military and Police Forces' health insurance were not affected by this change. This makes them a natural control group for our survival analysis, separating a general trend of annual change regarding the prospect of survival from the effect introduced by GES.

Since both databases share the anonymized patient ID codes, for each patient in the treatment database, we calculate their survival time as the total elapsed time between their first diagnosis and eventual appearance in the death database. If the cause of death corresponds to an unrelated ICD-10 code (neither principal nor related), then this is considered a right-censored case for our purposes. Similarly, if the patient does not appear in the death database, then they are considered surviving until the end of the study period (end of 2018) with right-censored death.

For the empirical survival analysis, we used the Kaplan–Meier estimator on different subsets of the patients' database, which correspond to relevant demographic subgroups, taking a confidence interval of 95%. We compared them using the log-rank test (21), for which we fixed a statistical significance level of 0.05.

To build a general survival model simultaneously encompassing the different demographic characteristics of the patients, we used the Cox proportional hazard model (22), treating categorical variables as dummies. We use the Akaike information criterion (23) to delineate the significant predictors from the variables mentioned above, but we force the inclusion of the variables corresponding to the year of diagnosis and the coverage by the GES guarantees to identify the effect of GES through the control group, as mentioned above.

All statistical analyses were programmed using Python 3.7 with the *lifelines* package for survival analysis.

# 3. Results

## 3.1 Trends in incidence and mortality

From Table 1, we observe that the annual number of CRC diagnoses increased by 58.6%, from 2,909 in 2009 to 4,615 in 2018. In the same period, the total number of CRC deaths increased by 53.6%, from 1,906 to 2,923. The mean age at diagnosis remained relatively constant at approximately 66 years old.

We computed crude and age-adjusted incidence and mortality rates to account for changes in the population (cases/100,000 inhabitants). Figure 2 presents the trends in prevalence, crude incidence, and mortality for the period under study. We observed significant increases in CRC crude prevalence, incidence, and mortality rates between 2009 and 2018. The prevalence rate increased 57% (from 62.4 to 97.7), the incidence rate increased 40% (from 20.5 to 28.7), and the crude mortality rate increased 38% (11.1 to 15.6).

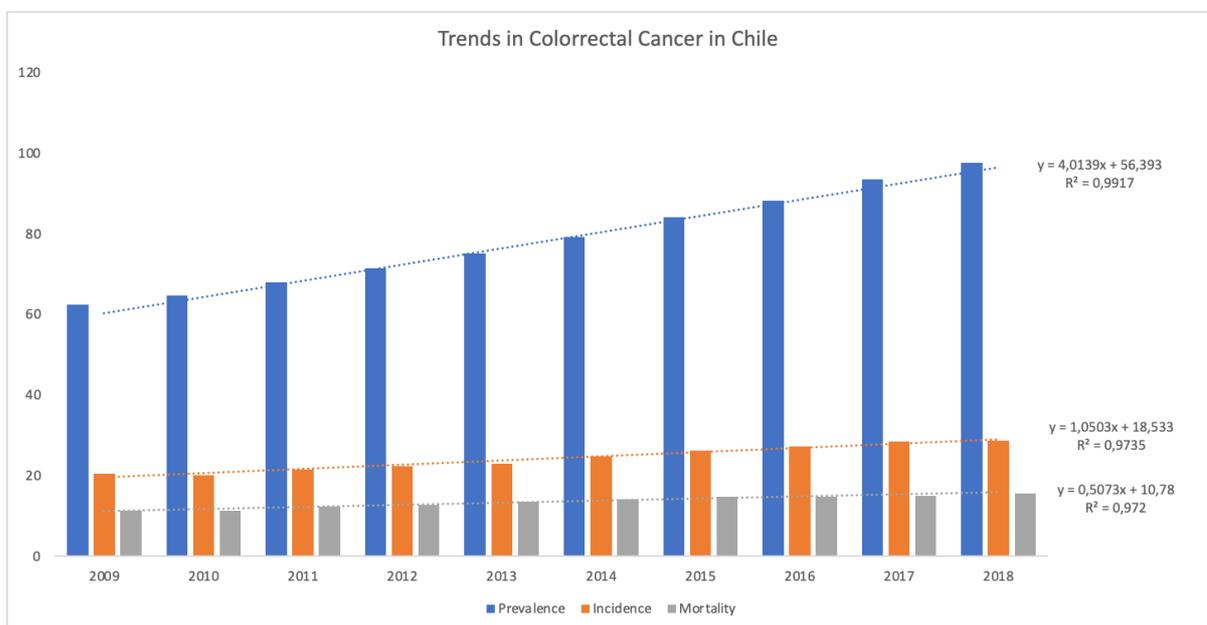

Figure 2. Crude prevalence, incidence, and mortality rates (cases/100,000 p.)

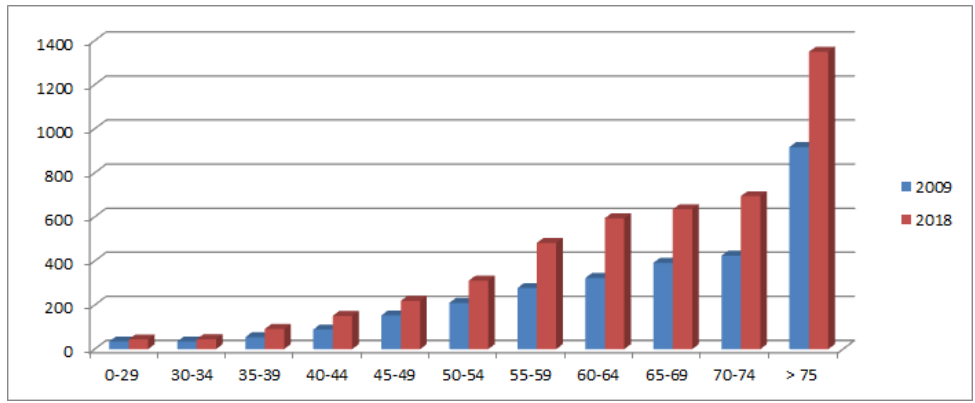

Figure 3. Crude incidence rates for different age groups, 2009 vs. 2018.

Appendix C presents yearly crude and age-standardized incidence and mortality rates for the period under study, 2009 to 2018. We include sex, insurance type, age group and region for the crude rates, while we construct the age-adjusted rates for men and women.

For crude incidence rates, we did not observe major differences between men and women. However, there was a larger increase in the crude incidence rates in men than in women (48% vs. 33%). We observe important differences in the incidence rates between those affiliated with FONASA public insurance vs. privately insured ISAPRES (average incidence rate 24,9 vs. 17,5 during the study period). For both groups, the increase in the crude incidence rate was similar (42% for FONASA and 40% for ISAPRE).

Most CRC patients were older than 55 years old. The relative increase in incidence rate was most pronounced in the age group of 40-44 years, followed by 35-39 years, 55-59 and 45-49 years (see Figure *3*).

Considerable differences were observed in the geographical distribution of crude incidence rates, with the lowest in the northern regions, increasing gradually toward the center of the country and reaching the most elevated values in the southern regions. During the study period, a constant increase in the crude incidence rate was detected in all regions, conserving geographical differences (Figure 4, left panel).

The age-standardized incidence rate increased from 15.7 to 18.6 (an 18% increase in the 2009-2018 period). We observe an increasing difference between the age-standardized rate of men and women, with 4.6 more cases/100,000 inhabitants in men than in women in 2018.

For crude mortality rates, again, we did not observe major differences between men and women. However, there was a larger increase in the crude mortality rates in men than in women (53% vs. 26%). For mortality, we also observe important differences considering the type of insurance (average mortality rate for FONASA 14.4 and for ISAPRE 5.9 during the study period). Moreover, the crude mortality rate increased by 100% for FONASA patients, while it only increased by 37% for the privately insured patients for the period under analysis.

Not surprisingly, CRC mortality rates increased with age. The increase in mortality rate was most pronounced in the 55-59 years age group, followed by 35-39 years and 45-49 years (59%, 56% and 37%, respectively).

Again, considerable differences were observed in the geographical distribution of crude mortality rates, with the lowest rate of 7.5 (northern region XV) and the largest rate of 14.9 (central region V). During the study period, an increase in crude mortality rate was detected in all regions, with similar geographical differences but greater variability (Figure 4, right panel). It is worth noting that region XIII (metropolitan region), with 40% of the population, had one of the lowest mortality rates that increased by only 26% during the period under study.

The age-standardized mortality rate increased 13% from 8.2 (2009) to 9.3 (2018). The age-standardized mortality rate was higher for men than women (11.4 vs. 5.9 average rates during the study period), with a higher increase in the mortality rate of males that increased 22%.

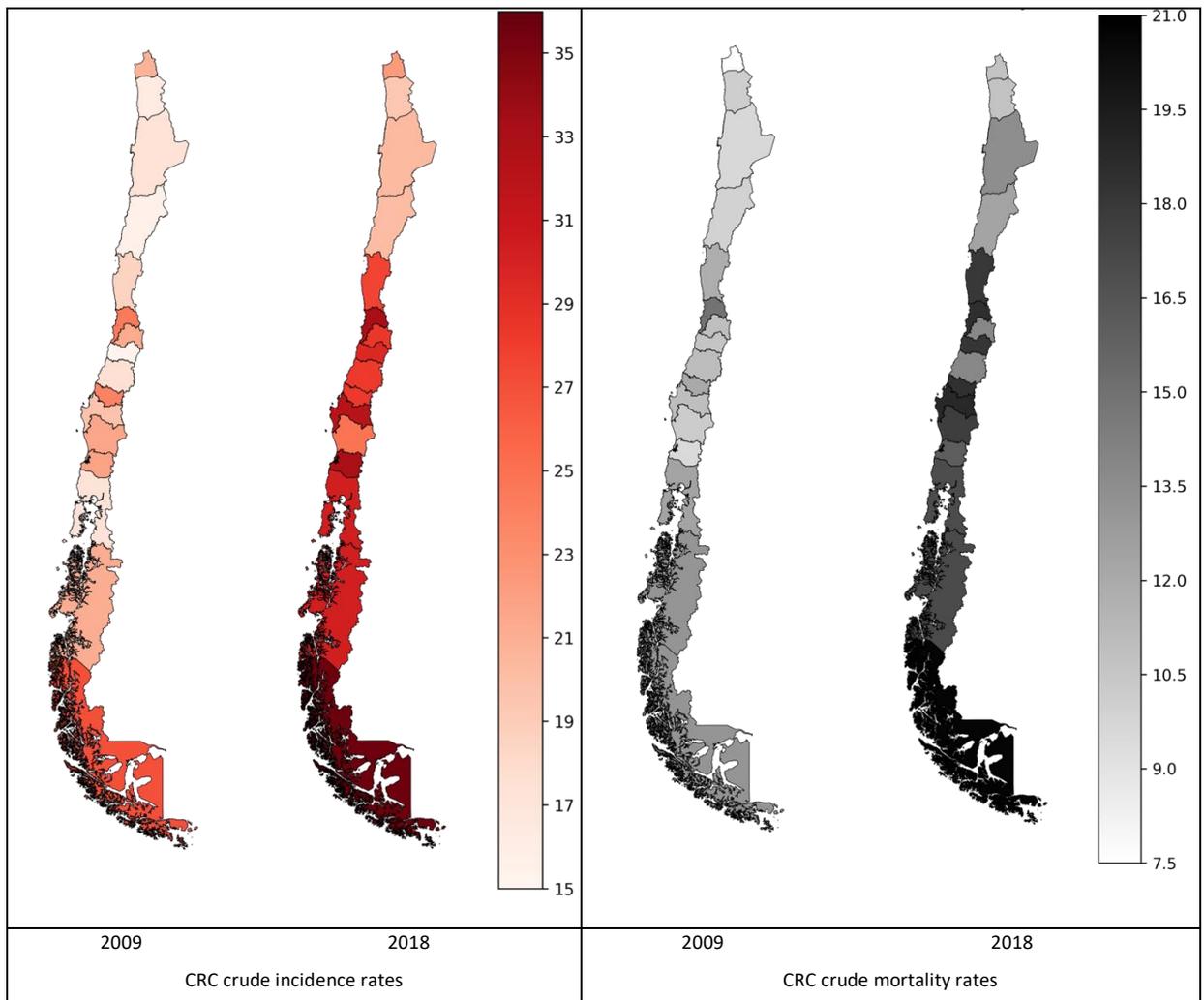

Figure 4. Comparison of regional crude rates (yearly number of cases/100,000 p.) between the start and the end of the study period. The left panel presents crude **incidence** rates, while the right panel presents the crude **mortality** rates.

## 3.2 Trends in survival rates

When analyzing empirical Kaplan–Meier survival rates, we observed an overall 51% five-year survival rate, considering all patients in the treatment database. If we constrain the limit to the first year, 27% of patients in the treatment database are no longer alive. Figure 5 shows the five-year Kaplan–Meier survival curve for all patients in the treatment database. The Kaplan–Meier curves did not show significant differences between men and women in the empirical survival rates, so we did not include the figure in the results.

Figure 6 shows the five-year Kaplan–Meier survival curves for patients in the treatment database, separated by age group. Not surprisingly, the survival rate decreases as age increases for patients older than 60 years old. Younger patients did not have a statistically significant difference in survival rates for either one- or five-year survival.

If we consider the cohort of patients using the year in which they were diagnosed, we can see an improvement in the survival rates. Figure 7 shows the five-year Kaplan–Meier survival curves for patients in the treatment database, separated by year of inclusion in the database. The curves are truncated because of the lack of follow-up for patients entering the database in later years.

Remarkably, a significant difference was observed between public (FONASA) and private health insurance systems (ISAPRE), with 47% and 68% five-year survival rates, respectively. We also observe differences within the public health insurance subgroups, showing a poorer outcome as the socioeconomic condition worsens (Group D: 54%, C: 52%, B: 46%, A: 39%). However, we noticed that the survival curves for patients with FONASA D and C were not significantly different. We also show that the patients in the control group (with military and police forces insurance) have survival rates that are halfway between ISAPRE and FONASA D patients.

If we compare the first year after the CRC diagnosis, only 13% of ISAPRE patients died, while this proportion was approximately 24% in both the FONASA C and D groups, increasing to 30 and 38% in the B and A groups with the poorest socioeconomic status (Figure 8).

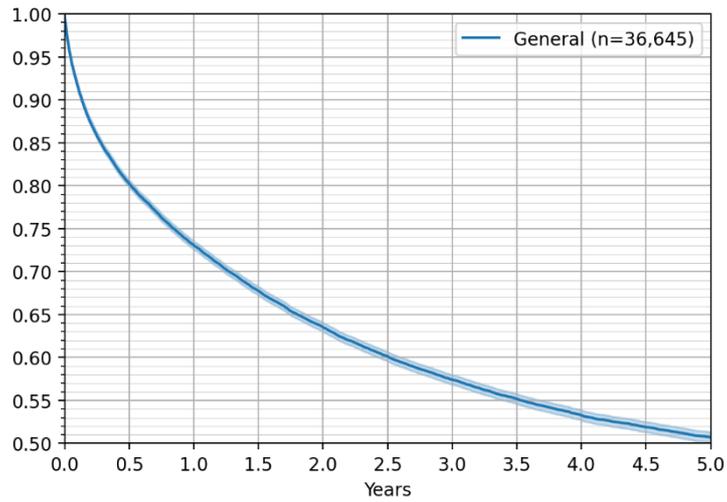

Figure 5. Five-year Kaplan–Meier survival curve for all patients in the treatment database, with 95% confidence interval.

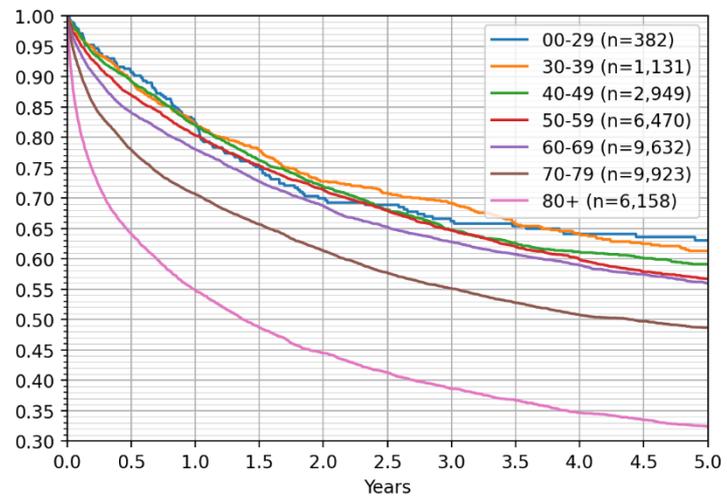

Figure 6. Five-year Kaplan–Meier survival curves for patients in the treatment database, separated by age groups (confidence intervals omitted for clarity).

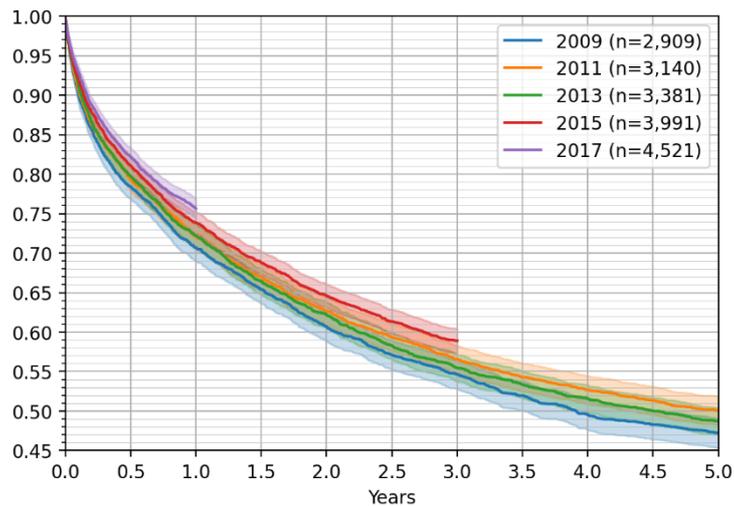

Figure 7. Five-year Kaplan–Meier survival curves for patients in the treatment database, separated by year of inclusion in the database, with 95% confidence intervals (only odd-numbered years are shown for clarity).

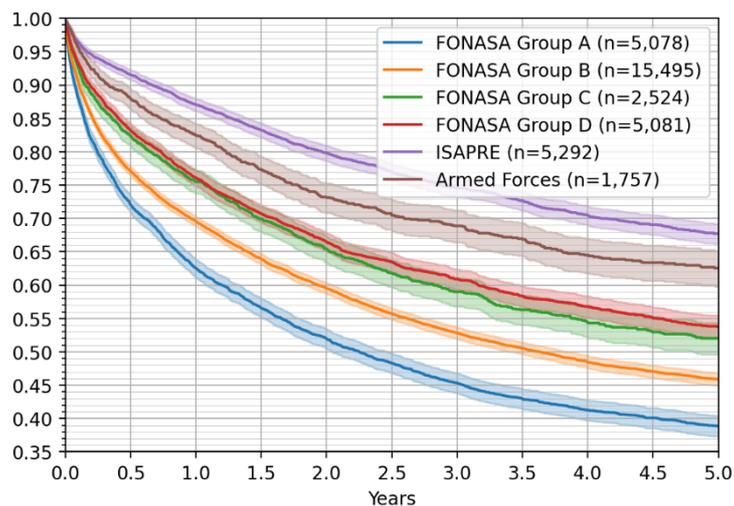

Figure 8. Five-year Kaplan–Meier survival curves for patients in the treatment database by health care insurance, with 95% confidence intervals.

We used Cox's proportional hazards regression model, or Cox model, to study the effect of different individual factors on the survival function. Factors considered include year of diagnosis, a dummy variable to classify whether the year of diagnosis belongs to the period when the GES plan was active, sex, type of tumor (colon or rectal cancer), insurance, age, and geographical region.

We consider as the base case of female patients, 70-74 years old, under FONASA B insurance group, living in the Metropolitan Region (XIII), with a colon tumor and entering the treatment database during the GES period. For this analysis, we grouped FONASA C and D patients. We also grouped some adjacent geographical regions to increase the number of patients in each group. The results are summarized in Table 2, where the first column contains the coefficient with its associated confidence interval, the second column shows the corresponding odds ratio at the coefficient's central value, and the third column shows the p value for the null hypothesis corresponding to equality of the base and the affected covariable.

Year of diagnosis has a negative coefficient, indicating that patients diagnosed with CRC in the latest years of the study period had greater five-year survival probabilities than those diagnosed at the beginning of the period (p value 0.013). GES period has no significant effect on survival.

Sex and type of tumor were dummy variables. We found that women had greater survival probabilities than did men (p value $< 5·10^{-5}$), but having a colon tumor had a small effect on decreasing survival probability compared with patients who had tumors in the rectum (p value 0.016).

For the insurance type, when comparing the base case – FONASA B – with all other groups, we found significantly different survival probabilities (p value $< 5·10^{-5}$). The control insurance, FONASA C+D and ISAPRE showed much higher survival probabilities compared with FONASA B, with a coefficient for the control group close to the ISAPRE insured patients. FONASA A had a remarkably lower survival probability than FONASA B.

Age is negatively correlated with the survival probabilities. Younger patients had a significantly higher survival probability, while older patients had a lower survival probability. The differences among all age groups were statistically significant (p value $<5·10^{-4}$). Moreover, the coefficients are ordered from smallest to largest (negative for younger than 70 and positive for older than 74).

Patients in the XIII region had a higher survival probability than patients located elsewhere in the country, which was statistically significant for regions II, V, VI, VII, VIII, IX, X, and XVI (p value $< 5·10^{-3}$) and regions I+XV, III+IV, and XI+XII (p value <0.05).

|  |  | Coefficient | Odds-ratio | p value |
|---|---|---|---|---|
|  | Year of diagnosis | -0.0151 ± 0.012 | 0.985 | 0.013 |
|  | With GES → Without GES | 0.0378 ± 0.0659 | 1.0385 | 0.261 |
| Sex | Female → Male | 0.13 ± 0.0345 | 1.1388 | $<5\cdot10^{-5}$ |
|  | Colon → Rectum | -0.0457 ± 0.0371 | 0.9553 | 0.0158 |
| INSURANCE | FONASA B → Control | -0.5866 ± 0.0971 | 0.5562 | $<5\cdot10^{-5}$ |
|  | FONASA B → FONASA A | 0.3193 ± 0.0479 | 1.3761 | $<5\cdot10^{-5}$ |
|  | FONASA B → FONASA C + D | -0.1794 ± 0.0466 | 0.8358 | $<5\cdot10^{-5}$ |
|  | FONASA B → ISAPRE | -0.6689 ± 0.0632 | 0.5123 | $<5\cdot10^{-5}$ |
| AGE | 70-74 → 00-29 | -0.5374 ± 0.2085 | 0.5843 | $<5\cdot10^{-5}$ |
|  | 70-74 → 30-34 | -0.5134 ± 0.2027 | 0.5985 | $<5\cdot10^{-5}$ |
|  | 70-74 → 35-39 | -0.4255 ± 0.1495 | 0.6534 | $<5\cdot10^{-5}$ |
|  | 70-74 → 40-44 | -0.318 ± 0.1192 | 0.7276 | $<5\cdot10^{-5}$ |
|  | 70-74 → 45-49 | -0.3743 ± 0.0983 | 0.6878 | $<5\cdot10^{-5}$ |
|  | 70-74 → 50-54 | 2-0.2926 ± 0.0827 | 0.7463 | $<5\cdot10^{-5}$ |
|  | 70-74 → 55-59 | -0.2643 ± 0.0744 | 0.7677 | $<5\cdot10^{-5}$ |
|  | 70-74 → 60-64 | -0.2567 ± 0.0705 | 0.7736 | $<5\cdot10^{-5}$ |
|  | 70-74 → 65-69 | -0.132 ± 0.0651 | 0.8764 | 0.0001 |
|  | 70-74 → 75-79 | 0.1497 ± 0.0634 | 1.1615 | $<5\cdot10^{-5}$ |
|  | 70-74 → 80-84 | 0.4309 ± 0.065 | 1.5386 | $<5\cdot10^{-5}$ |
|  | 70-74 → 85+ | 0.8227 ± 0.0669 | 2.2766 | $<5\cdot10^{-5}$ |
| REGION | XIII → I + XV | 0.1165 ± 0.1128 | 1.1236 | 0.0429 |
|  | XIII → II | 0.3348 ± 0.1028 | 1.3977 | $<5\cdot10^{-5}$ |
|  | XIII → III + IV | 0.1301 ± 0.0789 | 1.1389 | 0.0012 |
|  | XIII → V | 0.1048 ± 0.0556 | 1.1104 | 0.0002 |
|  | XIII → VI | 0.2046 ± 0.0835 | 1.227 | $<5\cdot10^{-5}$ |
|  | XIII → VII | 0.1492 ± 0.0797 | 1.1609 | 0.0002 |
|  | XIII → VIII | 0.1707 ± 0.0618 | 1.1861 | $<5\cdot10^{-5}$ |
|  | XIII → IX | 0.2438 ± 0.0757 | 1.276 | $<5\cdot10^{-5}$ |
|  | XIII → X | 0.2385 ± 0.0802 | 1.2693 | $<5\cdot10^{-5}$ |
|  | XIII → XI + XII | 0.1277 ± 0.1232 | 1.1362 | 0.0422 |
|  | XIII → XIV | 0.0998 ± 0.1033 | 1.1049 | 0.0583 |
|  | XIII → XVI | 0.2602 ± 0.0942 | 1.2972 | $<5\cdot10^{-5}$ |

*Table 2. Results for the Cox proportional hazards regression model.*

The impact of some of the observable characteristics on the estimated survival probabilities is presented in Figure 9, where we compare the Cox hazard curves for several cases against the base case (female patient with colon tumor, diagnosed during GES period and FONASA B insurance).

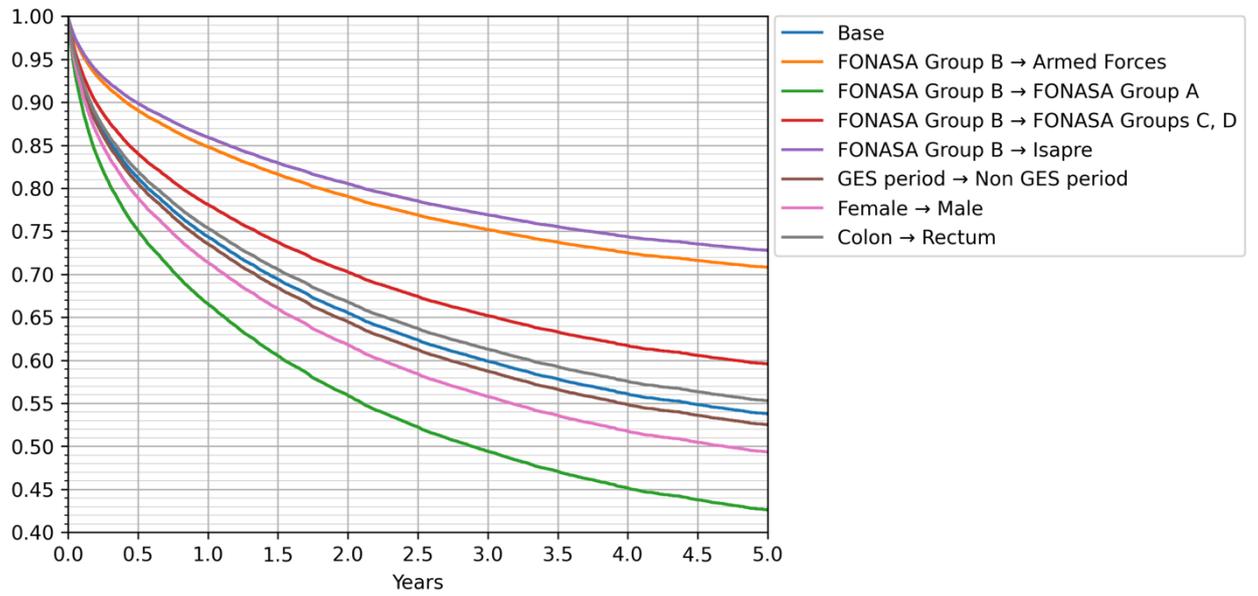

Figure 9: Cox proportional hazards regression model. Base: FONASA B, GES period, Female, Colon.

# 4. Discussion

## 4.1 Main findings

<u>Incidence, prevalence & mortality</u>

Previous national statistics estimated CRC incidence from the death registry and data on diagnosed cancer cases from 1998 to 2012 from four regional population registries (24). In our study, we used individual-level information crossing hospital discharge and death registries to estimate incidence, prevalence, and mortality rates. This allows for a more precise estimation of crude incidence without the need for extrapolations across time or regions.

Our results confirm a gradual, marked, and significant increase in CRC incidence (+40%), prevalence (+57%), and mortality (+38%) in the ten-year study period (2009-2018). It is noteworthy that the increase in the incidence and mortality was practically parallel in the study period, with only a slightly superior increase in incidence when compared to mortality. We found a *geographical gradient* with increasing incidence and mortality rates from the north to the south of the country. These geographical differences are somewhat parallel to the ethnic differences in the country (Mapuches in the South and European immigrants in extreme South region, Aymaras in the North and dominantly Eurasian in the central regions) (25).

When considering different *age groups*, most CRC patients were older than 55 years, but we also observed a marked incidence increase in the age group from 35 to 50 years (+55%), representing 10% of all incident CRC. This finding is of special interest when designing the starting age of screening programs. This aligns with recent literature reporting an increasing early onset of CRC in patients younger than 50 years old and similar rates of adenoma and advanced neoplasia in screening colonoscopy in individuals age 45 to 49 years as in individuals older than 50 years old (26) (27) (28) (29). Based on these experiences, updated recommendations suggest starting CRC screening at age 45 years (30).

The age-standardized incidence and mortality rates make it possible to compare Chile to other countries. CRC is strongly associated with a high human development index (HDI), probably related to sociodemographic, nutritional, and health care factors. Chile has one of the highest HDIs of the region, and thus, it is not surprising that the age-standardized incidence of CRC is also the highest compared to the rest of South America but significantly lower than European countries, Eastern Asia, Oceania and the United States (31) (32) (33) (3).

Survival and its dependence from different factors

According to the American Cancer Society, the average 5-year survival rate of CRC is 65%, and the stage at diagnosis is one of the most determinant factors in survival: in localized disease (stage I), the 5-year survival is over 90% compared to 71% in regional disease, worsening to 14% in metastatic disease (stage IV) (34) (35) (36) (37). We found a 50.7% general 5-year survival rate, inferior to the abovementioned data from the USA. Unfortunately, we could not obtain information on disease stage from the Chilean public registries used. However, we explored the impact of other observable factors, such as age, region, health insurance, and the GES program, on survival rates.

Age is clearly an important factor in survival. However, we only found significantly shorter survival in patients older than 70 years. This greater lethality can be explained by comorbidities, frequently found in this age group, although we did not analyze the causes of deaths unrelated to CRC. Additionally, a higher rate of new-onset cardiovascular disease has been observed in CRC survivors, which is even higher when the patient has received chemotherapy (38) and may have a larger impact in older patients.

We found a significantly better survival rate in the metropolitan region (XIII) than in the rest of the country. A plausible explanation in our statistics is known health access barriers, determined by the centralized distribution of resources, for example, the number of high-quality hospitals, a higher concentration of subspecialty physicians and access to prompt colonoscopy (39) (40). In fact, the Metropolitan Region has a physician rate of 212 physicians/100,000 people, which is almost twice as high as that in the rest of the country (119 in and 117 physicians per 100,000 people in the northern and southern regions, respectively) (40).

Using public databases, we were unable to identify socioeconomic, educational, and cultural differences, medical information on comorbidities, willingness to participate in screening programs or adherence to medical treatment. Nevertheless, patients' insurance can be used as a proxy for some of the above (19).

This social gradient has been reported in the literature from wealthy countries, with high life expectancy and adequate health insurance coverage, demonstrating that a low socioeconomic position is a risk factor for lower survival, partially explained by a more advanced stage at diagnosis (41) (42) (43) (44).

We found a significantly higher incidence in FONASA vs. ISAPRE. The difference between health insurances was even more striking when analyzing patient survival: 67.7% of patients affiliated with private health insurance (ISAPREs) were alive 5 years after CRC diagnosis, with a survival curve similar to that of developed countries (35) (36), which is significantly better than only 40% in the lowest income subgroup A of the public health insurance FONASA (see Figure 8).

Again, resource availability may play a role in the differences in CRC survival. In FONASA, there are 920 patients per physician, which is significantly higher than the 276 patients per physician working in the private sector (40). Additionally, ISAPRE patients have access to screening and preventive procedures included in most insurance plans, while FONASA patients can access such treatments only if there is a strong suspicion of CRC.

Moreover, there is probably a race and ethnicity factor in some populations. In a study using the U.S. National Cancer Database, Hao et al. confirmed that even with adequate health insurance, Black and Hispanic populations are still less likely to receive standard care for CRC (45). In Chile, there are significant differences in the type of health insurance and the ethnicity of the population, with FONASA being more indigenous than ISAPREs in comparison with the rest of the population (19).

In Chile, the GES program for CRC was implemented in 2014 to guarantee equity in prompt access to the best available treatment, with financial protection, once the patient is suspected to have CRC, independent of health care insurance. However, our study reveals that until 2018, survival rates had not significantly improved since GES implementation. The GES program has proven to slightly decrease mortality rates in cervix, breast, and gallbladder cancer (46). These diseases were included in the GES program in 2006 and either had a national screening program (Pap smear for women between 25 and 65 years old and mammography for women between 50 and 59 years old) or their prevention was part of the GES program itself (cholecystectomy for people between 35 and 49 years old with gallbladder stones).

The question remains whether CRC results are not visible because five years is not enough time to assess the impact of the GES program or whether early diagnosis for CRC is the dominant factor that can only be managed with screening programs that are not included in the CRC GES program.

## 4.2 Opportunities for the health policy

We believe there are three different levels to improve CRC results in Chile: prevention, early diagnosis, and prompt access to the optimal treatment.

Prevention aims to develop a healthy lifestyle, with balanced nutrition and regular physical activity, obesity prevention, and tobacco and alcohol use prevention. In Chile, there is a nationwide program promoting healthy lifestyles ([www.eligevivirsano.cl](www.eligevivirsano.cl)) but with no specific focus on preventing CRC.

Early diagnosis is only possible with screening programs. There are different alternatives for CRC screening programs with pros/cons and limitations. Lin et al. presented an overview of the evidence of the different screening strategies (47). There is evidence that there is a need for long-standing screening programs to see the impact on CRC reduction (3) (48), and therefore, any effort to implement a screening program should have a long-term perspective.

As mentioned above, in Chile, we have had some localized CRC screening programs, with a high rate of detection of adenomas during colonoscopy (36.5%) and early-detection of CRC (0.9%), allowing endoscopic resection in most cases (16) (49). The territorial limitation of these programs makes it unlikely to reduce national CRC incidence or mortality.

Prompt access to the optimal treatment was addressed by the introduction of the GES program for CRC. However, our results showed that this is probably insufficient to reduce the incidence and mortality of CRC.

### 4.3 Suggestions for the authorities

To reduce CRC mortality in Chile, we need a free access national screening program, including people over 45 years, followed by an enhanced version of the GES program, assuring quick access to endoscopic diagnostic and therapeutic procedures, surgical interventions, chemotherapy, and organized follow-up. Of course, this program requires a considerable increase in the capacity of endoscopic centers and all specialists involved in the diagnosis and treatment of CRC.

The creation of a national cancer registry with comprehensive patient information, including both public and private health care systems, is crucial to evaluate the effectiveness of this or any other program, helping to focus the use of public resources.

### 4.3 Study strengths and limitations

To our knowledge, this is the first study to compare and relate information from different registries. Our results provide objective data about the steady increase in incidence, prevalence and mortality of CRC and situates Chile in the highest range of South America, approaching developed countries. In addition, to our knowledge, lethality and survival were analyzed for the first time at the national level, demonstrating a clear dependence on regional and socioeconomic factors.

Our study has several limitations. First, our results depend on the limitations and correctness of national databases. Second, we could identify only patients with hospital admissions or those who died from CRC without hospital admission but not people who had only ambulatory colonoscopic resection of a lesion with "cancer in situ". Third, we do not know the stage of the disease at the time of diagnosis, a factor of utmost prognostic importance.

Therefore, as part of our ongoing research, we are developing mathematical and econometric models to estimate the distribution of CRC diagnoses among the different stages at the time of diagnosis. Furthermore, we also quantified the effect of different screening programs on the cancer stages at diagnosis as a potential explanation for the difference in survival rates observed between private and publicly insured patients.

## Conclusions

Our data and results show the real situation of CRC and its changes across the country in the period 2009–2018. Reliable individual patient information was obtained, allowing us to estimate the incidence and mortality of colorectal cancer in Chile. The factors influencing the survival of patients were analyzed. Our study provides evidence for the impact of socioeconomic inequalities on CRC incidence, mortality, and 5-year survival rates. Incidence and mortality rates were considerably lower, and the survival was significantly longer for patients enrolled in the private insurance system than for those enrolled in the public national health insurance system. Even within the publicly insured group, clear differences were observed in survival depending on socioeconomic conditions. Moreover, we could not find a significant effect of the GES program in improving the survival of CRC patients. Our work provides valuable information that helps to compare Chile with the rest of the world. It provides a solid basis for the construction of new health care programs with differential emphasis in different regions, which can truly improve the reality of CRC.

# Appendix A - FONASA groups

| | Income level (*) | Insurance coverage |
|---|---|---|
| Group A | - No income or immigrants<br>- Family subsidy (Law 18.020) | 100% coverage in the public network |
| Group B | - Up to 319.000 CLP monthly | 100% coverage in the public network<br>Access to limited private services with copay |
| Group C | - Between 319.000 and 465.740 CLP montly<br>- With three or more dependants qualify to level B | 90% coverage in the public network<br>Access to limited private services with copay |
| Group D | - More than 465.740 CLP montly<br>- With three or more dependants qualify to level C | 80% coverage in the public network<br>Access to limited private services with copay |

(*) 1 USD = 810 CLP

Table 3. Description of FONASA Groups.

# Appendix B - CRC ICD-10 diagnosis codes

### PRINCIPAL COLORECTAL CANCER ICD-10 DIAGNOSIS CODES & DESCRIPTIONS

| Code | Description | Code | Description |
|---|---|---|---|
| C180 | Malignant neoplasm of cecum | D120 | Benign neoplasm of cecum |
| C182 | Malignant neoplasm of ascending colon | D122 | Benign neoplasm of ascending colon |
| C183 | Malignant neoplasm of hepatic flexure | D123 | Benign neoplasm of transverse colon |
| C184 | Malignant neoplasm of transverse colon | D124 | Benign neoplasm of descending colon |
| C185 | Malignant neoplasm of splenic flexure | D125 | Benign neoplasm of sigmoid colon |
| C186 | Malignant neoplasm of descending colon | D126 | Benign neoplasm of colon, unspecified |
| C187 | Malignant neoplasm of sigmoid colon | D127 | Benign neoplasm of rectosigmoid junction |
| C189 | Malignant neoplasm of colon, unspecified | D128 | Benign neoplasm of rectum |
| C19X | Malignant neoplasm of rectosigmoid junction | D374 | Neoplasm of uncertain behavior of colon |
| C20X | Malignant neoplasm of rectum | D375 | Neoplasm of uncertain behavior of rectum |
| D010 | Carcinoma in situ of colon | | |
| D011 | Carcinoma in situ of rectosigmoid junction | | |
| D012 | Carcinoma in situ of rectum | | |

Table 4. Principal CRC ICD-10 diagnosis codes & descriptions.

### COLORECTAL CANCER RELATED ICD-10 DIAGNOSIS CODES & DESCRIPTIONS

| Code | Description | Code | Description |
|---|---|---|---|
| C181 | Malignant neoplasm of appendix | K566 | Intestinal adhesions [bands] with complete obstruction |
| C188 | Malignant neoplasm of overlapping sites of colon | K621 | Rectal polyp |
| C210 | Malignant neoplasm of anus, unspecified | K624 | Stenosis of anus and rectum |
| C211 | Malignant neoplasm of anal canal | K625 | Hemorrhage of anus and rectum |

| Code | Description | Code | Description |
|---|---|---|---|
| C218 | Malignant neoplasm of overlapping sites of rectum, anus and anal canal | K626 | Ulcer of anus and rectum |
| C227 | Other specified carcinomas of liver | K629 | Disease of anus and rectum, unspecified |
| C229 | Malignant neoplasm of liver, not specified as primary or secondary | K630 | Abscess of intestine |
| C260 | Malignant neoplasm of intestinal tract, part unspecified | K631 | Perforation of intestine (nontraumatic) |
| C268 | Malignant neoplasm of spleen | K639 | Disease of intestine, unspecified |
| C269 | Malignant neoplasm of ill-defined sites within the digestive system | K914 | Postprocedural complete intestinal obstruction |
| C480 | Malignant melanoma of skin, unspecified | K922 | Gastrointestinal hemorrhage, unspecified |
| C762 | Malignant neoplasm of endocrine gland, unspecified | R688 | Other general symptoms and signs |
| C785 | Malignant neoplasm of endocrine gland, unspecified | Z031 | Encounter for administrative examinations, unspecified |
| C786 | Malignant neoplasm of endocrine gland, unspecified | Z080 | Encounter for follow-up examination after completed treatment for malignant neoplasm |
| C787 | Malignant neoplasm of endocrine gland, unspecified | Z081 | Encounter for follow-up examination after completed treatment for malignant neoplasm |
| C809 | Malignant neoplasm associated with transplanted organ | Z082 | Encounter for follow-up examination after completed treatment for malignant neoplasm |
| C80X | Malignant neoplasm associated with transplanted organ | Z087 | Encounter for follow-up examination after completed treatment for malignant neoplasm |
| C97X | Malignant neoplasm of lymphoid, hematopoietic and related tissue, unspecified | Z088 | Encounter for follow-up examination after completed treatment for malignant neoplasm |
| D013 | Carcinoma in situ of anus and anal canal | Z089 | Encounter for follow-up examination after completed treatment for malignant neoplasm |
| D019 | Carcinoma in situ of digestive organ, unspecified | Z432 | Encounter for attention to ileostomy |
| D097 | Carcinoma in situ of thyroid and other endocrine glands | Z433 | Encounter for attention to colostomy |
| D377 | Benign neoplasm, unspecified site | Z510 | Encounter for antineoplastic radiation therapy |
| D379 | Benign neoplasm, unspecified site | Z511 | Encounter for antineoplastic radiation therapy |
| D489 | Neoplasm of uncertain behavior, unspecified | Z512 | Encounter for antineoplastic immunotherapy |
| D630 | Anemia in neoplastic disease | Z932 | Ileostomy status |
| K564 | Gallstone ileus | Z933 | Colostomy status |
| K565 | Other impaction of intestine | | |

Table 5. CRC related ICD-10 diagnosis codes & descriptions.

# Appendix C – Incidence and mortality rates

|  |  | 2009 | 2010 | 2011 | 2012 | 2013 | 2014 | 2015 | 2016 | 2017 | 2018 | Change 2018/2009 [%] |
|---|---|---|---|---|---|---|---|---|---|---|---|---|
| Crude incidence rates (yearly number of cases/100,000 p) | | | | | | | | | | | | |
| Total | | 20.5 | 20.2 | 21.6 | 22.4 | 22.9 | 24.9 | 26.2 | 27.2 | 28.5 | 28.7 | 40.0 |
| Gender | Men | 19.9 | 19.7 | 20.8 | 21.8 | 22.7 | 25.3 | 26.0 | 27.0 | 28.9 | 29.4 | 47.7 |
| | Women | 21.0 | 20.6 | 22.2 | 23.1 | 23.2 | 24.6 | 26.5 | 27.4 | 28.1 | 28.0 | 33.3 |
| Insurance | Fonasa | 20.8 | 20.5 | 21.8 | 22.5 | 23.1 | 25.3 | 27.3 | 28.5 | 29.6 | 29.5 | 41.8 |
| | Isapre | 15.9 | 14.7 | 14.3 | 16.8 | 15.4 | 18.1 | 18.8 | 17.9 | 20.9 | 22.2 | 39.6 |
| Age group | 00-29 | 0.5 | 0.5 | 0.5 | 0.5 | 0.5 | 0.6 | 0.5 | 0.4 | 0.5 | 0.6 | 20.0 |
| | 30-34 | 2.9 | 2.7 | 3.2 | 3.0 | 3.7 | 3.2 | 2.7 | 3.2 | 4.0 | 3.3 | 13.8 |
| | 35-39 | 4.7 | 4.4 | 4.9 | 6.3 | 5.0 | 5.4 | 6.7 | 6.3 | 7.2 | 7.0 | 48.9 |
| | 40-44 | 7.3 | 6.6 | 9.3 | 9.4 | 9.4 | 10.4 | 9.2 | 9.5 | 11.9 | 11.8 | 61.6 |
| | 45-49 | 14.3 | 12.9 | 14.0 | 14.7 | 14.7 | 14.4 | 17.5 | 18.1 | 21.0 | 18.7 | 30.8 |
| | 50-54 | 22.8 | 22.6 | 21.8 | 25.3 | 26.1 | 28.2 | 27.4 | 29.6 | 31.8 | 27.8 | 21.9 |
| | 55-59 | 35.5 | 33.0 | 36.8 | 36.5 | 40.2 | 42.8 | 45.3 | 44.0 | 49.0 | 47.5 | 33.8 |
| | 60-64 | 55.8 | 56.6 | 62.9 | 57.2 | 59.6 | 55.3 | 61.1 | 63.4 | 68.1 | 69.5 | 24.6 |
| | 65-69 | 88.7 | 83.6 | 82.9 | 98.6 | 96.6 | 98.2 | 98.9 | 104.2 | 103.3 | 94.7 | 6.8 |
| | 70-74 | 122.9 | 118.4 | 124.5 | 128.7 | 122.2 | 144.9 | 136.8 | 148.8 | 142.8 | 148.4 | 20.7 |
| | 75-79 | 158.4 | 158.3 | 155.3 | 162.6 | 150.3 | 155.5 | 182.2 | 173.1 | 171.4 | 183.2 | 15.9 |
| | 80-84 | 200.8 | 187.9 | 201.0 | 190.4 | 195.8 | 204.9 | 213.3 | 201.4 | 199.9 | 205.6 | 2.4 |
| | 85-89 | 246.3 | 237.0 | 256.9 | 195.4 | 199.9 | 258.0 | 229.5 | 245.5 | 235.4 | 237.1 | -3.8 |
| | ≥ 90 | 264.4 | 253.9 | 192.7 | 237.7 | 237.3 | 250.1 | 264.2 | 221.3 | 215.0 | 211.7 | -19.8 |
| Region | I | 16.3 | 12.9 | 12.6 | 14.9 | 17.2 | 16.5 | 16.8 | 17.6 | 19.0 | 19.4 | 19.0 |
| | II | 17.3 | 17.6 | 17.0 | 19.7 | 18.8 | 17.8 | 18.5 | 19.1 | 22.6 | 20.3 | 17.3 |
| | III | 15.6 | 11.5 | 13.5 | 14.0 | 15.3 | 18.8 | 18.7 | 15.5 | 20.0 | 20.1 | 28.8 |
| | IV | 18.6 | 18.5 | 20.5 | 23.8 | 19.1 | 22.1 | 23.6 | 26.4 | 26.0 | 27.6 | 48.4 |
| | V | 24.4 | 27.9 | 29.5 | 27.2 | 28.0 | 27.6 | 32.1 | 32.6 | 34.9 | 32.9 | 34.8 |
| | VI | 15.1 | 17.9 | 19.1 | 19.7 | 20.2 | 22.4 | 22.3 | 29.2 | 32.3 | 29.5 | 95.4 |
| | VII | 17.6 | 12.8 | 17.6 | 19.4 | 18.3 | 24.1 | 25.0 | 29.1 | 26.2 | 28.2 | 60.2 |
| | VIII | 19.6 | 18.7 | 22.5 | 22.1 | 24.5 | 24.4 | 25.5 | 29.9 | 28.5 | 32.0 | 63.3 |
| | IX | 21.5 | 22.3 | 19.3 | 13.9 | 24.9 | 28.4 | 24.6 | 26.6 | 27.5 | 24.8 | 15.3 |
| | X | 17.2 | 16.8 | 24.9 | 20.6 | 20.1 | 23.9 | 21.6 | 24.1 | 26.8 | 30.2 | 75.6 |
| | XI | 21.2 | 18.0 | 19.8 | 28.4 | 22.4 | 28.0 | 32.7 | 28.7 | 25.6 | 30.2 | 42.5 |
| | XII | 26.9 | 24.8 | 34.5 | 25.6 | 46.0 | 30.0 | 31.0 | 28.9 | 28.6 | 35.7 | 32.7 |
| | XIII | 21.2 | 21.1 | 21.0 | 23.1 | 22.8 | 25.5 | 27.5 | 26.8 | 28.1 | 28.3 | 33.5 |
| | XIV | 21.9 | 19.4 | 23.4 | 43.7 | 27.2 | 27.1 | 24.6 | 25.5 | 29.6 | 32.9 | 50.2 |
| | XV | 20.7 | 19.5 | 16.0 | 14.4 | 20.4 | 27.2 | 31.6 | 26.9 | 34.2 | 22.3 | 7.7 |
| | XVI | 23.8 | 17.8 | 22.7 | 22.5 | 24.4 | 23.5 | 28.4 | 27.4 | 32.2 | 28.2 | 18.5 |
| Age-standardized incidence rates (yearly number of cases/100,000 p) | | | | | | | | | | | | |
| Total | | 15.7 | 15.2 | 15.9 | 16.4 | 16.4 | 17.3 | 17.8 | 18.2 | 18.9 | 18.6 | 18.5 |
| Gender | Men | 17.5 | 16.9 | 17.4 | 17.9 | 18.2 | 19.8 | 19.8 | 20.1 | 21.2 | 21.1 | 20.6 |
| | Women | 14.4 | 13.9 | 14.8 | 15.2 | 15.0 | 15.4 | 16.2 | 16.8 | 17.2 | 16.5 | 14.6 |

Table 6. Yearly crude incidence rates, including gender, insurance, age group and region. Total yearly age standardized incidence rates and by gender.

| | | 2009 | 2010 | 2011 | 2012 | 2013 | 2014 | 2015 | 2016 | 2017 | 2018 | Change 2018/2009 [%] |
|---|---|---|---|---|---|---|---|---|---|---|---|---|
| Crude mortality rates (yearly number of cases/100,000 p) | | | | | | | | | | | | |
| Total | | 11.3 | 11.4 | 12.3 | 12.8 | 13.5 | 14.1 | 14.8 | 14.8 | 15.1 | 15.6 | 38.1 |
| Gender | Men | 10.5 | 11.5 | 11.9 | 12.5 | 13.1 | 14.1 | 14.8 | 14.6 | 15.0 | 16.1 | 53.3 |
| | Women | 12.0 | 11.4 | 12.7 | 13.2 | 13.8 | 14.1 | 14.9 | 15.0 | 15.1 | 15.1 | 25.8 |
| Insurance | Fonasa | 12.1 | 12.1 | 12.6 | 13.6 | 13.9 | 15.0 | 15.9 | 15.9 | 16.3 | 16.6 | 37.2 |
| | Isapre | 3.9 | 4.5 | 5.2 | 5.0 | 6.0 | 6.3 | 6.9 | 6.3 | 7.2 | 7.8 | 100 |
| Age group | 00-29 | 0.1 | 0.3 | 0.2 | 0.1 | 0.2 | 0.2 | 0.2 | 0.1 | 0.2 | 0.1 | 0 |
| | 30-34 | 1.7 | 0.8 | 0.9 | 0.9 | 1.3 | 0.8 | 0.8 | 1.6 | 0.7 | 1.0 | -41.2 |
| | 35-39 | 1.8 | 1.0 | 1.6 | 1.9 | 2.0 | 2.0 | 2.5 | 1.9 | 2.5 | 2.8 | 55.6 |
| | 40-44 | 2.5 | 2.5 | 2.9 | 2.8 | 4.0 | 3.7 | 2.7 | 3.2 | 4.3 | 3.4 | 36.0 |
| | 45-49 | 5.1 | 5.7 | 5.2 | 5.4 | 6.8 | 6.1 | 7.6 | 6.8 | 7.7 | 7.0 | 37.3 |
| | 50-54 | 9.8 | 8.8 | 8.5 | 10.1 | 12.2 | 9.4 | 11.6 | 11.2 | 12.4 | 11.5 | 17.3 |
| | 55-59 | 13.0 | 14.9 | 15.3 | 15.7 | 17.8 | 19.2 | 20.8 | 17.4 | 18.7 | 20.7 | 59.2 |
| | 60-64 | 24.9 | 26.1 | 27.0 | 25.0 | 29.6 | 27.7 | 27.5 | 24.5 | 26.7 | 29.6 | 18.9 |
| | 65-69 | 45.4 | 41.8 | 45.0 | 48.9 | 46.8 | 43.8 | 47.5 | 50.1 | 43.9 | 45.8 | 0.9 |
| | 70-74 | 65.6 | 71.1 | 76.0 | 76.6 | 68.4 | 80.5 | 76.4 | 75.9 | 75.3 | 76.8 | 17.1 |
| | 75-79 | 100.2 | 93.2 | 104.4 | 107.6 | 102.2 | 101.6 | 102.5 | 106.8 | 95.0 | 99.4 | -0,8 |
| | 80-84 | 139.7 | 139.5 | 157.9 | 145.5 | 148.6 | 161.5 | 156.1 | 157.6 | 153.8 | 156.4 | 12.0 |
| | 85-89 | 204.9 | 189.8 | 201.5 | 187.0 | 185.9 | 226.7 | 220.5 | 213.9 | 230.9 | 215.9 | 5.4 |
| | ≥ 90 | 264.4 | 258.1 | 186.9 | 270.3 | 259.2 | 234.5 | 274.2 | 246.1 | 227.0 | 229.3 | -13.3 |
| Region | I | 10.1 | 7.8 | 6.3 | 10.4 | 8.3 | 7.2 | 9.7 | 10.5 | 9.4 | 10.7 | 5.9 |
| | II | 9.6 | 8.7 | 8.9 | 12.1 | 9.4 | 11.7 | 10.9 | 12.9 | 13.5 | 13.5 | 40.6 |
| | III | 9.9 | 5.6 | 10.0 | 6.2 | 9.5 | 13.1 | 9.3 | 10.3 | 10.2 | 12.3 | 24.2 |
| | IV | 11.8 | 10.5 | 11.1 | 12.3 | 13.3 | 13.4 | 13.8 | 16.9 | 17.0 | 18.0 | 52.5 |
| | V | 14.9 | 15.3 | 16.3 | 16.6 | 16.3 | 16.0 | 20.6 | 18.5 | 16.5 | 18.5 | 24.2 |
| | VI | 10.6 | 10.1 | 12.0 | 11.4 | 13.8 | 15.3 | 14.2 | 16.8 | 19.0 | 18.1 | 70.8 |
| | VII | 11.0 | 9.5 | 11.4 | 13.0 | 12.2 | 12.6 | 13.3 | 17.0 | 15.4 | 13.8 | 25.5 |
| | VIII | 10.9 | 11.9 | 13.5 | 13.1 | 13.5 | 14.4 | 14.5 | 15.2 | 15.2 | 18.8 | 72.5 |
| | IX | 10.2 | 13.7 | 12.6 | 13.1 | 15.8 | 16.5 | 16.5 | 16.6 | 17.7 | 17.7 | 73.5 |
| | X | 12.3 | 9.9 | 12.8 | 15.6 | 14.6 | 14.6 | 15.5 | 11.6 | 15.6 | 16.9 | 37.4 |
| | XI | 13.1 | 13.0 | 11.9 | 19.6 | 11.7 | 10.6 | 13.5 | 18.2 | 15.2 | 17.0 | 29.8 |
| | XII | 13.2 | 13.0 | 20.9 | 17.7 | 15.7 | 21.6 | 14.3 | 18.9 | 20.4 | 20.7 | 56.8 |
| | XIII | 10.9 | 11.2 | 11.8 | 11.9 | 12.9 | 13.3 | 13.9 | 13.4 | 13.4 | 13.8 | 26.7 |
| | XIV | 9.5 | 13.1 | 10.9 | 15.5 | 15.7 | 15.3 | 17.3 | 15.7 | 17.3 | 16.0 | 68.4 |
| | XV | 7.5 | 8.4 | 6.9 | 10.4 | 12.0 | 20.6 | 15.6 | 9.0 | 18.6 | 10.8 | 44.0 |
| | XVI | 12.0 | 13.0 | 12.9 | 12.4 | 17.3 | 17.8 | 19.3 | 19.1 | 20.2 | 18.4 | 53.3 |
| Age-standardized mortality rates (yearly number of cases/100,000 p) | | | | | | | | | | | | |
| Total | | 8.2 | 8.2 | 8.5 | 8.7 | 9.0 | 9.1 | 9.3 | 9.1 | 9.1 | 9.3 | 13.4 |

| Gender | Men | 9.1 | 9.7 | 9.7 | 9.9 | 10.2 | 10.6 | 10.9 | 10.5 | 10.5 | 11.1 | 22.0 |
|---|---|---|---|---|---|---|---|---|---|---|---|---|
| | Women | 7.6 | 7.1 | 7.6 | 7.9 | 8.1 | 7.9 | 8.0 | 8.0 | 8.1 | 7.9 | 3.9 |

Table 7. Yearly crude mortality rates, including gender, insurance, age group and region. Total yearly age standardized mortality rates and by gender.

# Appendix D – Kaplan-Meier survival curves

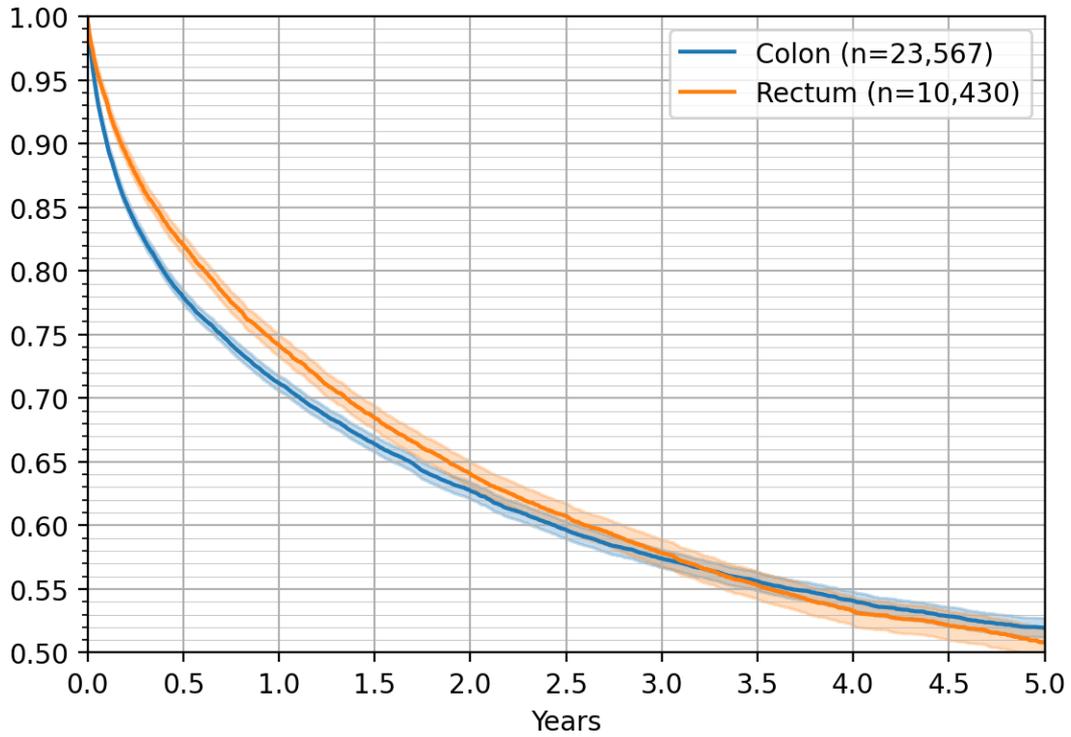

Figure 10. Five-year Kaplan-Meier survival curves considering location of the tumor, for patients in the treatment database.

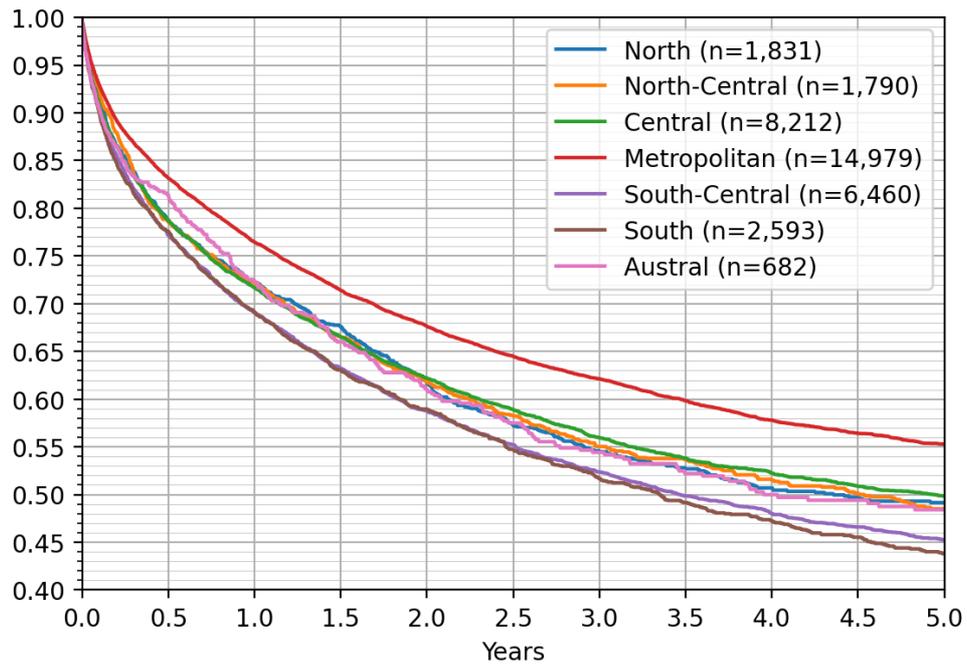

Figure 11. Five-year Kaplan-Meier survival curves for patients in the treatment database, separated by regional macrozones.